\documentclass[aps,reprint,superscriptaddress,shortbibliography,prl]{revtex4-1}

\usepackage{graphicx, amsmath, verbatim, dsfont, amsfonts, color}
\usepackage{braket}
\usepackage[us]{datetime}

\begin{document}

\title{Anomalous Low-Temperature enhancement of Supercurrent in Topological-Insulator Nanoribbon Josephson Junctions: Evidence for Low-Energy Andreev Bound States}

\author{Morteza~Kayyalha}
\email{mkayyalh@purdue.edu}
\affiliation{School of Electrical and Computer Engineering and Birck Nanotechnology Center, Purdue University, West Lafayette, Indiana 47907, USA}
\author{Mehdi~Kargarian}
\affiliation{Department of Physics, Condensed Matter Theory Center and Joint Quantum Institute, University of Maryland, College Park, Maryland 20742, USA}
\author{Aleksandr~Kazakov}
\affiliation{Department of Physics and Astronomy, Purdue University, West Lafayette, Indiana 47907, USA}
\author{Ireneusz~Miotkowski}
\affiliation{Department of Physics and Astronomy, Purdue University, West Lafayette, Inidana 47907, USA}
\author{Victor~M.~Galitski}
\affiliation{Department of Physics, Condensed Matter Theory Center and Joint Quantum Institute, University of Maryland, College Park, Maryland 20742, USA}
\author{Victor~M.~Yakovenko}
\affiliation{Department of Physics, Condensed Matter Theory Center and Joint Quantum Institute, University of Maryland, College Park, Maryland 20742, USA}
\author{Leonid~P.~Rokhinson}
\affiliation{Department of Physics and Astronomy, Purdue University, West Lafayette, Indiana 47907, USA}
\affiliation{School of Electrical and Computer Engineering and Birck Nanotechnology Center, Purdue University, West Lafayette, Indiana 47907, USA}
\author{Yong~P.~Chen}
\email{yongchen@purdue.edu}
\affiliation{Department of Physics and Astronomy, Purdue University, West Lafayette, Indiana 47907, USA}
\affiliation{School of Electrical and Computer Engineering and Birck Nanotechnology Center, Purdue University, West Lafayette, Indiana 47907, USA}
\affiliation{Purdue Quantum Center, Purdue University, West Lafayette, Indiana 47907, USA}
\affiliation{WPI-AIMR International Research Center on Materials Sciences, Tohoku University, Sendai 980-8577, Japan}

\begin{abstract}
We report anomalous enhancement of the critical current at low temperatures in gate-tunable Josephson junctions made from topological insulator BiSbTeSe$_2$ nanoribbons with superconducting Nb electrodes. In contrast to conventional junctions, as a function of the decreasing temperature $T$, the increasing critical current $I_c$ exhibits a sharp upturn at a temperature $T_*$ around 20$\%$ of the junction critical temperatures for several different samples and various gate voltages. The $I_c$ vs. $T$ demonstrates a short junction behavior for $T>T_*$, but crosses over to a long junction behavior for $T<T_*$ with an exponential $T$-dependence $I_c \propto \exp\big(-k_B T/\delta \big)$, where $k_B$ is the Boltzmann constant. The extracted characteristic energy-scale $\delta$ is found to be an order of magnitude smaller than the induced superconducting gap of the junction. We attribute the long-junction behavior with such a small $\delta$ to low-energy Andreev bound states (ABS) arising from winding of the electronic wavefunction around the circumference of the topological insulator nanoribbon (TINR). Our TINR-based Josephson junctions with low-energy ABS are promising for future topologically protected devices that may host exotic phenomena such as Majorana fermions.
\end{abstract}

\maketitle

Three-dimensional (3D) topological insulators (TI) are characterized by insulating bulk and non-trivial conducting surface states, where the spin is helically locked perpendicular to the momentum, and the carriers are massless Dirac fermions with linear energy-momentum dispersion \cite{Hasan2010,Qi2011,Hasan2010a}. Theoretical work by Fu and Kane \cite{Fu2008} has predicted that, once coupled to an s-wave superconductor, the surface states of TI's can undergo unconventional superconducting pairing, which can provide a useful platform to study exotic phenomena such as topological superconductivity and Majorana fermions \cite{Qi2011,Fu2008}. In contrast to the conventional spin-singlet superconductivity, the induced superconductivity in the surface states of a 3D TI \cite{Fu2008} is a mixture of singlet and triplet pairings due to the lifted spin degeneracy \cite{Gorkov2001,Tkachov2013,Gong2017}. Furthermore, Andreev bound states (ABS) formed within a superconductor-TI-superconductor (S-TI-S) Josephson junction (JJ) can exhibit a robust zero-energy crossing when the phase difference between the two superconductors is $\pi$, giving rise to Majorana modes \cite{Fu2008, Tkachov2013}. Possible probes of topological superconductors/junctions may include the tunneling spectroscopy, the current-phase relation (CPR), and temperature dependence of the critical current \cite{Beenakker2013a,Kwon2004,Olund2012,Burset2014,Ghaemi2016,Tkachov2017}.

In recent years, S-TI-S JJ's with two- and three-dimensional TI's have been extensively studied. Gate-tunable supercurrent and Josephson effects, such as Fraunhofer patterns and Shapiro steps, have also been observed \cite{Sacepe2011,Zhang2011,Veldhorst2012,Williams2012,Qu2012,Yang2012,Sochnikov2013,Oostinga2013,Cho2013,Kurter2013,Finck2014,Lee2014a,Wiedenmann2015,Stehno2016,Bocquillon2017,Jauregui2018}. However, in many of the devices studied so far, the bulk of the TI can have notable contributions to the transport properties of the junction and make it difficult to separate out the contribution of the surface states.

In this work, we use the topological insulator BiSbTeSe$_2$ with a distinct advantage that at low temperatures the bulk is insulating and only the surface states contribute to electrical transport \cite{Xu2014,Xu2016,Jauregui2018}. We obtain nanoribons of BiSbTeSe$_2$ using the exfoliation technique and fabricate superconductor-(TI nanoribon)-superconductor (S-TINR-S) JJ's. Due to the enhanced surface to volume ratio, uniform cross-sectional area, and relatively small size, TINR-based devices have shown to be an excellent platform to study topological transport, exhibiting ballistic conduction and $\pi$-Berry-phase Aharonov-Bohm effects \cite{Hong2014,Cho2015,Jauregui2016}, and are also predicted to be promising for the study of topological superconductivity \cite{Cook2011,Ilan2014}. In our TINR-based JJ's, in contrast to conventional junctions, we observe a sharp upturn of the critical current $I_c$ for temperatures $T$ below $\sim20\%$ of the junction critical temperature $T_c$. Interestingly, this upturn temperature ($\sim 0.2T_c$) is observed in a variety of JJ's with different gate voltages $V_g$'s. We interpret the experimental results using a phenomenological model for junctions based on TINR's. This model relates the enhancement of $I_c$ at low temperatures to the ABS whose energy scale is around an order of magnitude smaller than the induced superconducting gap. The reduced energy scale of the ABS is attributed to the winding of their wavefunction around the circumference of the TINR. Such ABS are in the long junction limit and give rise to an exponential enhancement of $I_c$ with decreasing $T$. Furthermore, we observe a sinusoidal current-phase relation (CPR) measured using an asymmetric superconducting quantum interference device (SQUID), consistent with the expectation for these samples at our measurement temperature.
 
High-quality single crystals of BiSbTeSe$_2$ were grown by the Bridgman technique \cite{Xu2014}. Flakes exfoliated out of our BiSbTeSe$_2$ crystals exhibit the ambipolar field effect, half-integer quantum Hall effect, and $\pi$ Berry's phase characteristic of the spin-helical Dirac fermion topological surface states (TSS) \cite{Xu2014,Xu2016}. We obtain BiSbTeSe$_2$ nanoribbons \cite{Jauregui2018} using the scotch-tape exfoliation technique and transfer them onto 300-nm-thick SiO$_2$/500-$\mu$m-thick highly-doped Si substrates, which are used as back gates. Nanoribbons of various width $W$ and thickness $t$ are then located using an optical microscope. Subsequently, electron beam lithography is performed to define two closely separated electrodes with a separation $L <$ 100 nm. Finally, a thin layer of Niobium (Nb) as a superconductor, 50-nm thick, is deposited in a DC sputtering system. Prior to Nb deposition, brief ($\sim3$ seconds) Ar ion milling is performed to improve the quality of Nb contacts to TINR's. We have previously observed large $I_c R_N$ product (where $R_N$ is the normal-state resistance) and multiple Andreev reflections in such TINR JJ's  \cite{Jauregui2018}, demonstrating the high quality of the junctions including the Nb-TINR interface. Inset of Fig. 1b depicts an atomic force microscope (AFM) image of a representative S-TINR-S junction (sample 1). We have studied a variety of TINR JJ's with electrode separation $L \sim 40-70$ nm, width $W \sim 250-400$ nm, and thickness $t \sim 30-50$ nm. These dimensions are measured by an AFM. Detailed parameters for all the samples studied are listed in Table S1 in the supplemental information (SI) \cite{SIPRL2017}.

Fig. 1a shows the ambipolar field effect in the two-terminal resistance $R$ vs. $V_g$ measured in sample 1 at $T=14.5$ K, above the superconducting critical temperature of Nb. By varying $V_g$, the carrier type in the TINR can be changed from n-type to p-type, and the chemical potential can be tuned into the bulk bandgap to be in the TSS. The gate voltage where the maximum of $R$ vs. $V_g$ occurs represents the charge neutrality point (CNP) which is $V_{CNP} \sim -15$ V for this sample.

\begin{figure*}\label{fig:1}
\centering\includegraphics[width=1.9\columnwidth]{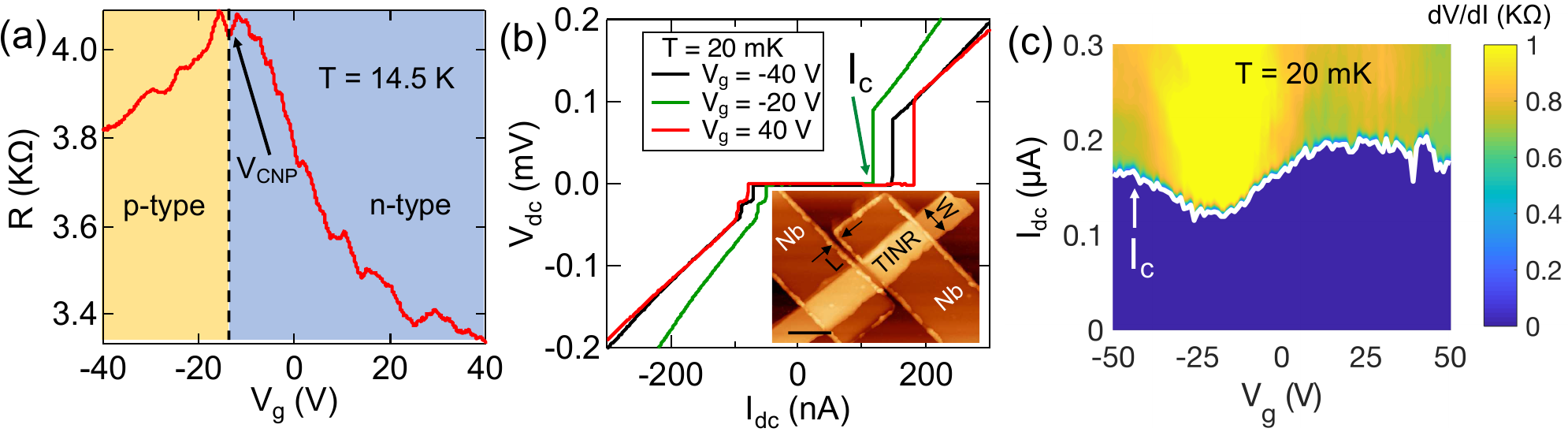}
\vspace{0in}
\caption{(a) Two-terminal $R$ vs. $V_g$ measured at $T = 14.5$ K, above the critical temperature $T_c^{Nb} = 7.5$ K of the Nb electrodes. Shaded regions highlight n and p doping of TINR. (b) The DC voltage $V_{dc}$ vs. the DC current $I_{dc}$ of the junction for different $V_g$'s at $T = 20$ mK. Inset: Atomic force microscope (AFM) image of sample 1 (from which all data in this figure are measured), a topological insulator (BiSbTeSe$_2$) nanoribbon (TINR)-based Josephson device with superconducting Nb electrodes. Scale bar is 0.5 $\mu m$. (c) Color map of the two-terminal $dV/dI$ vs. $V_g$ and $I_{dc}$ at $T = 20$ mK. An AC excitation current $I_{ac}$ = 1 nA was used for the $dV/dI$ measurement. Solid white line marks the junction critical current $I_c$ vs. $V_g$. Data in (b-c) is measured by sweeping $I_{dc}$ from -300 nA to 300 nA at various $V_g$'s.}
\vspace{0in} 
\end{figure*}

The junction critical temperature ($T_c$ $\sim$ 0.5 - 2.2 K), the temperature below which the junction resistance vanishes, is much lower than the critical temperature of Nb ($T_c^{Nb}$ $\sim$ 7.5 K) in our S-TINR-S junctions. The DC voltage $V_{dc}$ vs. the DC current $I_{dc}$, measured in sample 1 when sweeping $I_{dc}$ from -300 nA to 300 nA at $T$ = 20 mK for a few different $V_g$'s is plotted in Fig. 1b. When $I_{dc}$ is small, the voltage across the junction is zero, indicating that the junction is in its superconducting state and supports a supercurrent ($I_{dc}$). However, once the current is increased above some critical current (defined as $I_c$, marked by the arrow for the $V_g$ = -20 V curve), the junction leaves the superconducting state and transitions to the normal state with a finite voltage drop. Fig. 1c shows the color map of the two-terminal differential resistance $dV/dI$ vs. $V_g$ and $I_{dc}$ (swept from 0 to 300 nA) at $T$ = 20 mK. The solid white line in this figure marks the critical current $I_c$ of the junction. Notably, we observe that $I_c$ exhibits an ambipolar field effect (which has not been realized in previous devices  \cite{Cho2013,Kurter2013,Jauregui2018}) and reaches a minimum of $\sim 120$ nA near $V_{CNP} \sim -15$ V, consistent with the peak in $R$ vs. $V_g$ measurement (Fig. 1a).

Fig. 2a shows the $T$-dependence of $I_c$ for three different $V_g$'s in sample 1. Starting from $T_c$, $I_c$ increases with decreasing $T$. Notably, we observe an anomaly in $I_c$ vs. $T$ at an upturn temperature ($T_{*}\sim$ 0.36 K marked for the $V_g$ = 45 V dataset with $T_c \sim$ 2.2 K as an example), below which $I_c$ increases sharply and eventually reaches its largest value $I_c^{max}$ at the lowest accessible temperature ($T\sim20$ mK). The normalized critical current $I_c/I_c^{max}$ vs. the normalized temperature $T/T_c$ for this sample is depicted in Fig. 2b. Interestingly, $T_{*}$ is always $\sim0.2T_c$ for this sample regardless of the applied $V_g$. Fig. 2c plots $I_c/I_c^{max}$ vs. $T/T_c$ for five different samples, with each sample measured at a few $V_g$'s. We observe that $T_{*}/T_c$ remains $\sim0.2$ for all our TINR-based JJ's, regardless of their $T_c$ and $V_g$ (see Table S1 in the SI \cite{SIPRL2017}). Noteworthy, we observe an exponential enhancement of $I_c$ with decreasing $T$ for $T<T_*$ as highlighted by the solid red lines in Fig. 2b and c.

\begin{figure*}
\centering\includegraphics[width=1.9\columnwidth]{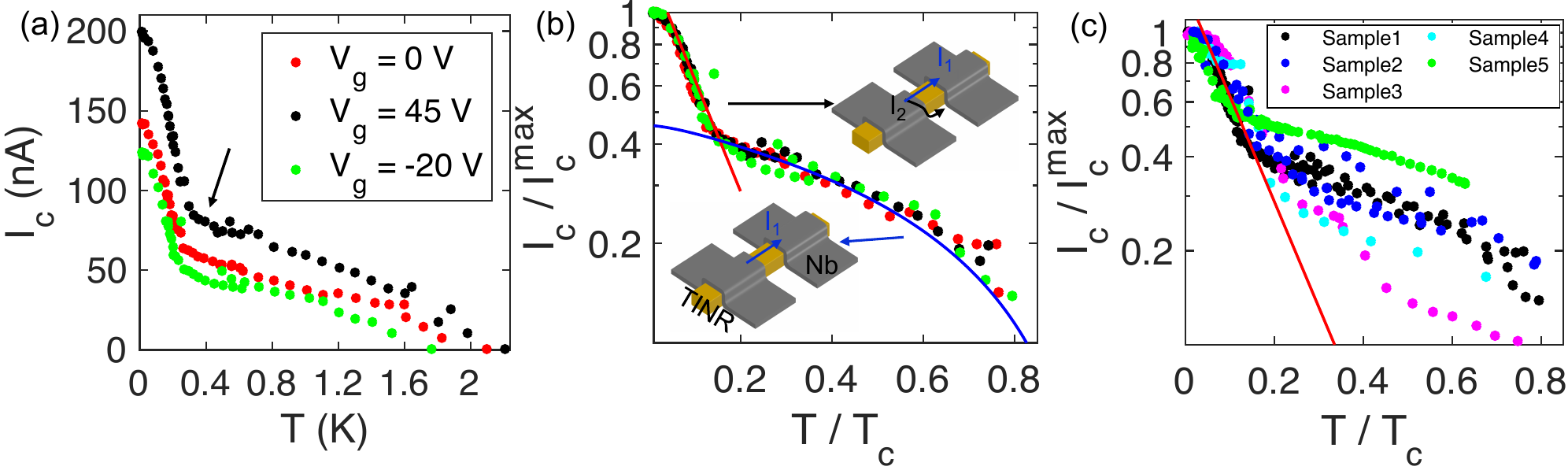}
\vspace{0in}
\caption{(a) Temperature dependence of $I_c$ for different $V_g$'s for sample 1. (b) Normalized $I_c/I_c^{max}$ vs. normalized $T/T_c$ for data in (a) in log-linear scale. The solid blue line is the normalized $I_{c1}/I_{c1}^{max}$ (Eq. \ref{eq:2}) divided by factor 2.2 and the solid red line is a fit to $\exp (- \frac {k_B T}{\delta})$ with $\delta \sim 0.08\Delta$. The symbols have the same legends as in (a). Inset: cartoons of the TINR JJ depicting the current $I_1$ corresponding to the modes on the top surface and the current $I_2$ corresponding to the modes that extend around the circumference and flow through the bottom surface. Due to the exponential decay of $I_2$ with increasing $T$, only $I_1$ contributes to the critical current at high temperatures. (c) $I_c/I_c^{max}$ vs. $T/T_c$ in a log-linear scale for five different TINR-based Josephson devices measured at a few (1-3) $V_g$'s for each device. The exponential fit and the experimental data in (b) are also included in this plot as the solid red line and black symbols, respectively.}
\label{fig:2}
\vspace{0in} 
\end{figure*}

The anomalous temperature dependence of $I_c$ observed in our samples is radically different from that of conventional JJ's. While the $T$-dependence of our $I_c$ for $T_*<T<T_c$ maybe described by the behavior of a TI-based \textit{short} junction (e.g., solid blue line in Fig. 2b, as discussed more in our model presented below), $I_c$ of such short junctions is \textit{not} expected to exhibit any exponential behavior before it saturates at low temperatures \cite{Olund2012,Tkachov2013}. On the other hand, for \textit{long} junctions it has been demonstrated that $I_c$ increases exponentially with decreasing temperature \cite{Dubos2001,Golubov2004,Angers2008,Ke2016,Borzenets2016} before its eventual saturation at the low temperature limit. Therefore, the increase in $I_c$ vs. decreasing $T$ for $T_*<T<T_c$ followed by an exponential enhancement of $I_c$ for $T<T_*$ as observed in Fig. 2b suggests that $I_c$ in our samples may be dominated by a short junction behavior for $T>T_*$ and a long junction behavior for $T<T_*$. Such a transition from short to long junction behaviors may be related to the nature of the TSS in the TINR. Because, the TSS extend over the entire circumference of the TINR, the superconducting transport is carried by modes on both the top (corresponding to $I_1$ depicted in the inset of Fig. 2b) and bottom (corresponding to $I_2$ depicted in the inset of Fig. 2b) surfaces of the TINR, i.e., the total supercurrent $I=I_1+I_2$. 

For the TINR with a circumference $C=2W+2t$, the transverse momentum $k_y$, perpendicular to the current, is quantized as $k_y=\frac{2\pi}{C}(n+1/2)$, where $n$ is an integer \cite{Zhang2010,Bardarson2010}. Also note in our TINR the current flows between the superconducting contacts fabricated on the top surface. Therefore, the modes with $k_y$ near zero remain on the top surface and contribute to $I_1$, while the modes with $|k_y| \gg 0$ extend around the perimeter of the TINR and contribute to $I_2$. We note that the $k_y=0$ mode is prohibited in the TINR. 

The modes (corresponding to $I_1$) on the top surface travel a short distance $L$, the separation between the two Nb contacts, and are supposedly in the short-junction limit. We found our experimental data of $I_c$ vs. $T$ for $T>T_*$ can be described using the temperature-dependent supercurrent calculated for a ballistic short junction \cite{Golubov2004,Olund2012,Tkachov2013}, given by:
\begin{equation} \label{eq:1}
I_1(\phi,T) = N_1 \frac{e \pi \Delta(T)}{h} \sin(\frac{\phi}{2}) \tanh \big(\frac{\Delta(T) \cos \big(\frac{\phi}{2} \big)}{2k_BT} \big),
\end{equation}
where $h$ is the Plank constant, $k_B$ is the Boltzmann constant, $e$ is the electron charge, $N_1$ is the number of modes in the top surface, $\phi$ is the phase difference between the two superconductors, and $\Delta(T)$ is the induced superconducting gap. We assume a BCS temperature dependence for $\Delta(T)$ with $\Delta(T=0)=\Delta_0=1.76k_B T_c$ \cite{Tinkham2004}. We obtain the critical current $I_{c1} (T)$ by maximizing $I_1(\phi,T)$ over $\phi$ as:
\begin{equation} \label{eq:2}
I_{c1} (T) = \max _{\phi} \Big( I_1(\phi,T) \Big).
\end{equation}
We have plotted $I_{c1}(T)$ calculated from Eq. (\ref{eq:2}) to obtain the solid blue curve in Fig. 2b. The computed $I_{c1}(T)/I_{c1}^{max}$, where $I_{c1}^{max}=I_{c1}(T=0)$, is divided by 2.2 in order to show its agreement with experimental results for $T>T_*$ in the normalized version of $I_c/I_c^{max}$ in Fig. 2b (this indicates $I_c^{max} \sim 2.2 I_{c1}^{max}$, or $I_{c1}$ on the top surface contributes nearly half of the total $I_c$ at the low temperature limit).

In contrast, the modes (corresponding to $I_2$) flowing through the bottom surface extend over the entire circumference ($C\sim$ 700 nm for sample 1 shown in Fig. 2a and b) of the TINR (through the side surface) and hence travel a longer distance $d$ ($d \geq C \gg L$). We assume such modes are in the ballistic long-junction limit with $d \geq \xi$, where $\xi=\hbar v_F/\Delta \sim$ 640 nm is the superconducting coherence length of the junction and $v_F$ is the Fermi velocity. As a result, we observe a reduced energy gap $\delta =\hbar v_F/2\pi d$ for these modes \cite{Bardeen1972,Svidzinsky1973,Bagwell1992,Golubov2004,Borzenets2016}. In the limit of $T_{sat}<T<T_*$, where $T_{sat} \ll \delta/k_B$ is the temperature below which $I_c$ saturates, the critical current of these modes exhibits an exponential dependence on $T$, i.e. $I_c \propto \exp (-k_B T/\delta )$ \cite{Bardeen1972,Svidzinsky1973,Bagwell1992,Golubov2004,Borzenets2016}. This exponential dependence is seen in the experimental data in Fig. 2b. To extract $\delta$, we perform an exponential fit to $I_c$ for $T_{sat}<T<T_*$ (where we take $T_{sat} \sim 0.04T_c$) as depicted by the solid red line in Fig. 2b. The fit gives $\delta \sim 0.08\Delta$, corresponding to $d = \frac{\hbar v_F}{2\pi\delta} \sim$ 1.2 $\mu$m, which is quite close to $\sim \xi+C$. We have found similar trends for the extracted $d \sim C+\xi$ in other samples shown in Fig. 2c (see also Fig. S2d \cite{SIPRL2017}). We suggest that when the effective length $d$ is on the order of $\xi$, the extracted $\delta$ should be proportional to $1/(C + \xi)$ rather than $1/C$ (thus $d$ should be closer to $C+\xi$ rather than $C$). This reduced $\delta$ is also consistent with the discussions in Ref. \cite{Borzenets2016}. To highlight the influence of $\xi$ on $\delta$ (and $T_*$), we have plotted $\delta$ and $T_*$ vs. $1/(C + \xi)$ in Fig. S2b \cite{SIPRL2017}. Consistent with our expectation, we observe that $\delta$ (and $T_*$) increases with increasing $1/(C + \xi)$.

We can extract $N_1 \sim$ 1-5 for different samples from the fit of $I_{c1}$ as determined by Eq. (\ref{eq:2}) to the experimental results. The extracted value of $N_1$ is much smaller than the estimated total number of modes $N=k_F C/2\pi \sim$ 24-114, where $k_F= \sqrt{4 \pi \frac{C_g}{e} (V_g-V_{CNP})}$ is the Fermi wave vector and $C_g$= 12 nF/cm$^2$ is the parallel plate capacitance per unit area of a 300-nm SiO$_2$. Furthermore, we can estimate the number of modes $N_2$ corresponding to $I_2$ as $N_2=N-N_1 \sim (10-20)N_1$. This suggests that the majority of the modes in our TINRs are going around the circumference and through the bottom surface to contribute to $I_2$, consistent with the expectation that only modes with $k_y$ near zero contribute to $I_1$. We note that $I_c$ at the lowest $T$ is proportional to the number of modes and the energy scale of the ABS in both the long and short junction limits (i.e. the low-$T$ $I_1$ and $I_2$ are proportional to $N_1 \Delta_0$ and $N_2 \delta$, respectively). The extracted large $N_2 \sim (10-20) N_1$ and the small $\delta \sim 0.1\Delta_0$ imply that the contribution of $I_1$ and $I_2$ to the total critical current at low $T$ should be comparable, which is consistent with our experimental observations in Fig. 2b and c. For instance, $I_{c1}$ represented by the solid blue line in Fig. 2b approaches $\sim 50 \%$ of the total $I_c$ when extrapolated to the lowest $T$.

In the above phenomenological model, we have used one effective reduced gap $\delta$ to describe all the modes flowing around the circumference and through the bottom surface. However, in reality these modes can have different gaps depending on how far they travel between the two superconductors. Currently there is no theory for the temperature dependence of $I_c$ specific to TINR (considering the wrapping of the electronic wavefunction around the circumference). Further studies are required to fully understand the nature of the induced superconductivity in this system.

We have measured the CPR (supercurrent $I$ vs. phase $\phi$) in our TINR junction at $T$ = 20 mK using an asymmetric SQUID based on our TINR junction in parallel with a reference junction \cite{DellaRocca2007,Zgirski2011,SIPRL2017}. Fig. 3a depicts a scanning electron microscope (SEM) image of the SQUID. The measured CPR (symbols) is shown in Fig. 3b alongside a sinusoidal function (black curve), which describes well the measured CPR. 

It has been predicted that in a TI flake, regardless of the barrier height $Z$ imposed by a non-magnetic impurity, $k_y$ = 0 mode will have a transmission probability $D$ = 1 and will give rise to a highly skewed CPR  \cite{Tkachov2013}. However, in the TINRs, the $k_y$ = 0 mode is strictly prohibited. Effectively, the small transverse size of the TINR generates a gap in the TSS spectrum, making the system more sensitive to disorder and rendering the CPR more sinusoidal. For $k_y \neq$ 0, $D$ depends on $Z$ and is not necessarily 1, thus CPR is not necessarily highly skewed. Furthermore, in our SQUID-based measurement, we need to ensure that the current through the reference junction (part of our SQUID device) is sufficient to drive it to the normal state. Since $I_c$ of the reference junction is on the order of 10 to 20 $\mu$A (compared to $I_c \sim$ 20-200 nA in the TINRs), the electron temperature in the CPR measurement could be substantially larger compared to that in the $V_{dc}$-$I_{dc}$ measurements (used to extract $I_c$). Additionally, our TINRs are very sensitive to temperature and show a strong asymmetry between their critical current $I_c$ and return current $I_r$ (see Fig. S3 and SI for more details \cite{SIPRL2017}) due to the Joule heating (caused by $I_{dc}$). Overall, the sensitivity to disorder for modes with $k_y \neq$ 0 as well as the increased electron temperature due to the large $I_c$ of the reference junction may result in sinusoidal CPR in our TINR-based JJ's measured in the SQUID setup.

\begin{figure}
\centering\includegraphics[width=0.9\columnwidth]{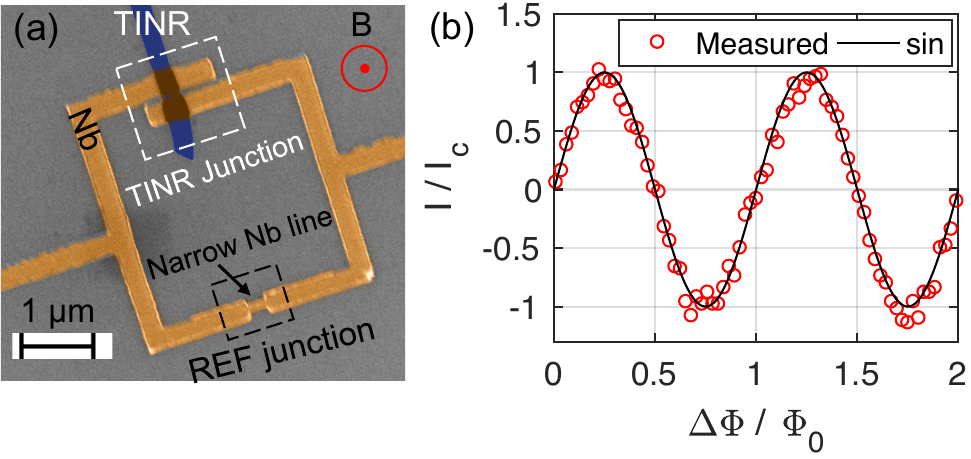}
\vspace{-0.2in}
\caption{(a) False-colored scanning electron microscope image of an asymmetric SQUID used to measure the current-phase relations (CPR) in our TINR-based JJ's. (b) Normalized current $I/I_c$ vs. normalized flux $\Delta \Phi /\Phi_0$, where $\Phi_0 = h/2e$ is the flux quanta, at $V_g$ = 20 V and $T$ = 20 mK. As the absolute value of the flux inside the superconducting SQUID is unknown, the experimental curve is shifted along the horizontal axis for comparison with a sinusoidal function.}
\label{fig:3}
\vspace{-0.2in} 
\end{figure}

In this paper, we present transport measurements of the JJ's based on nanoribbons of the bulk-insulating topological insulators BiSbTeSe$_2$ with superconducting Nb contacts. We experimentally find an anomalous behavior in the T-dependence of $I_c$ in a variety of junctions with different $T_c$ and $V_g$'s. For all samples, $I_c$ increases with decreasing temperature from $T_c$ to an upturn temperature ($\sim$ 0.2$T_c$), followed by an exponential increase with further decrease of the temperature. To understand our results, we introduce a phenomenological model based on the winding of the ABS around the circumference of the TINR. Our model relates the enhancement of $I_c$ at low temperatures to the anomalously small energy scale of ABS in the long-junction limit. Furthermore, our measured CPR shows a sinusoidal behavior, consistent with the expectation for such long JJ's under the experimental conditions. Our experimental observations indicate that our TINR junctions can be promising platforms for further exploration of topological superconductivity and Majorana fermions predicted in such systems \cite{Fu2008}.

\begin{acknowledgments}
M.K., L.P.R., and Y.P.C. acknowledge support from National Science Foundation (NSF) under Award DMR-1410942. M.K. and Y.P.C. also acknowledge partial support from NSF under Award EFMA-1641101. L.P.R. and A.K. also acknowledge partial support from the U.S. Department of Energy (DOE), Office of Basic Energy Sciences (BES) under Award DE-SC0008630 (L.P.R.) and Department of Defense Office of Naval Research Grant No. N000141410339 (A.K.). M.K. (UMD) acknowledges support by NSF DMR-1613029 and V.M.G. acknowledges support from DOE, BES (DESC0001911) and the Simons Foundation. The authors also acknowledge helpful discussions with Sergei Khlebnikov, Erhai Zhao, and Pouyan Ghaemi.
\end{acknowledgments}

\bibliographystyle{apsrev4-1}
\bibliography{biblography}

\begin{thebibliography}{50}%
\makeatletter
\providecommand \@ifxundefined [1]{%
 \@ifx{#1\undefined}
}%
\providecommand \@ifnum [1]{%
 \ifnum #1\expandafter \@firstoftwo
 \else \expandafter \@secondoftwo
 \fi
}%
\providecommand \@ifx [1]{%
 \ifx #1\expandafter \@firstoftwo
 \else \expandafter \@secondoftwo
 \fi
}%
\providecommand \natexlab [1]{#1}%
\providecommand \enquote  [1]{``#1''}%
\providecommand \bibnamefont  [1]{#1}%
\providecommand \bibfnamefont [1]{#1}%
\providecommand \citenamefont [1]{#1}%
\providecommand \href@noop [0]{\@secondoftwo}%
\providecommand \href [0]{\begingroup \@sanitize@url \@href}%
\providecommand \@href[1]{\@@startlink{#1}\@@href}%
\providecommand \@@href[1]{\endgroup#1\@@endlink}%
\providecommand \@sanitize@url [0]{\catcode `\\12\catcode `\$12\catcode
  `\&12\catcode `\#12\catcode `\^12\catcode `\_12\catcode `\%12\relax}%
\providecommand \@@startlink[1]{}%
\providecommand \@@endlink[0]{}%
\providecommand \url  [0]{\begingroup\@sanitize@url \@url }%
\providecommand \@url [1]{\endgroup\@href {#1}{\urlprefix }}%
\providecommand \urlprefix  [0]{URL }%
\providecommand \Eprint [0]{\href }%
\providecommand \doibase [0]{http://dx.doi.org/}%
\providecommand \selectlanguage [0]{\@gobble}%
\providecommand \bibinfo  [0]{\@secondoftwo}%
\providecommand \bibfield  [0]{\@secondoftwo}%
\providecommand \translation [1]{[#1]}%
\providecommand \BibitemOpen [0]{}%
\providecommand \bibitemStop [0]{}%
\providecommand \bibitemNoStop [0]{.\EOS\space}%
\providecommand \EOS [0]{\spacefactor3000\relax}%
\providecommand \BibitemShut  [1]{\csname bibitem#1\endcsname}%
\let\auto@bib@innerbib\@empty
\bibitem [{\citenamefont {Hasan}\ and\ \citenamefont {Kane}(2010)}]{Hasan2010}%
  \BibitemOpen
  \bibfield  {author} {\bibinfo {author} {\bibfnamefont {M.~Z.}\ \bibnamefont
  {Hasan}}\ and\ \bibinfo {author} {\bibfnamefont {C.~L.}\ \bibnamefont
  {Kane}},\ }\href {\doibase 10.1103/RevModPhys.82.3045} {\bibfield  {journal}
  {\bibinfo  {journal} {Reviews of Modern Physics}\ }\textbf {\bibinfo {volume}
  {82}},\ \bibinfo {pages} {3045} (\bibinfo {year} {2010})}\BibitemShut
  {NoStop}%
\bibitem [{\citenamefont {Qi}\ and\ \citenamefont {Zhang}(2011)}]{Qi2011}%
  \BibitemOpen
  \bibfield  {author} {\bibinfo {author} {\bibfnamefont {X.-L.}\ \bibnamefont
  {Qi}}\ and\ \bibinfo {author} {\bibfnamefont {S.-C.}\ \bibnamefont {Zhang}},\
  }\href {\doibase 10.1103/RevModPhys.83.1057} {\bibfield  {journal} {\bibinfo
  {journal} {Reviews of Modern Physics}\ }\textbf {\bibinfo {volume} {83}},\
  \bibinfo {pages} {1057} (\bibinfo {year} {2011})}\BibitemShut {NoStop}%
\bibitem [{\citenamefont {Hasan}\ and\ \citenamefont
  {Moore}(2010)}]{Hasan2010a}%
  \BibitemOpen
  \bibfield  {author} {\bibinfo {author} {\bibfnamefont {M.~Z.}\ \bibnamefont
  {Hasan}}\ and\ \bibinfo {author} {\bibfnamefont {J.~E.}\ \bibnamefont
  {Moore}},\ }\href {\doibase 10.1146/annurev-conmatphys-062910-140432}
  {\bibfield  {journal} {\bibinfo  {journal} {Annual Review of Condensed Matter
  Physics}\ }\textbf {\bibinfo {volume} {2}},\ \bibinfo {pages} {55} (\bibinfo
  {year} {2010})}\BibitemShut {NoStop}%
\bibitem [{\citenamefont {Fu}\ and\ \citenamefont {Kane}(2008)}]{Fu2008}%
  \BibitemOpen
  \bibfield  {author} {\bibinfo {author} {\bibfnamefont {L.}~\bibnamefont
  {Fu}}\ and\ \bibinfo {author} {\bibfnamefont {C.~L.}\ \bibnamefont {Kane}},\
  }\href {\doibase 10.1103/PhysRevLett.100.096407} {\bibfield  {journal}
  {\bibinfo  {journal} {Physical Review Letters}\ }\textbf {\bibinfo {volume}
  {100}},\ \bibinfo {pages} {096407} (\bibinfo {year} {2008})}\BibitemShut
  {NoStop}%
\bibitem [{\citenamefont {{Gor'kov}}\ and\ \citenamefont
  {Rashba}(2001)}]{Gorkov2001}%
  \BibitemOpen
  \bibfield  {author} {\bibinfo {author} {\bibfnamefont {L.~P.}\ \bibnamefont
  {{Gor'kov}}}\ and\ \bibinfo {author} {\bibfnamefont {E.~I.}\ \bibnamefont
  {Rashba}},\ }\href {\doibase 10.1103/PhysRevLett.87.037004} {\bibfield
  {journal} {\bibinfo  {journal} {Physical Review Letters}\ }\textbf {\bibinfo
  {volume} {87}},\ \bibinfo {pages} {037004} (\bibinfo {year}
  {2001})}\BibitemShut {NoStop}%
\bibitem [{\citenamefont {Tkachov}\ and\ \citenamefont
  {Hankiewicz}(2013)}]{Tkachov2013}%
  \BibitemOpen
  \bibfield  {author} {\bibinfo {author} {\bibfnamefont {G.}~\bibnamefont
  {Tkachov}}\ and\ \bibinfo {author} {\bibfnamefont {E.~M.}\ \bibnamefont
  {Hankiewicz}},\ }\href {\doibase 10.1103/PhysRevB.88.075401} {\bibfield
  {journal} {\bibinfo  {journal} {Physical Review B}\ }\textbf {\bibinfo
  {volume} {88}},\ \bibinfo {pages} {075401} (\bibinfo {year}
  {2013})}\BibitemShut {NoStop}%
\bibitem [{\citenamefont {Gong}\ \emph {et~al.}(2017)\citenamefont {Gong},
  \citenamefont {Kargarian}, \citenamefont {Stern}, \citenamefont {Yue},
  \citenamefont {Zhou}, \citenamefont {Jin}, \citenamefont {Galitski},
  \citenamefont {Yakovenko},\ and\ \citenamefont {Xia}}]{Gong2017}%
  \BibitemOpen
  \bibfield  {author} {\bibinfo {author} {\bibfnamefont {X.}~\bibnamefont
  {Gong}}, \bibinfo {author} {\bibfnamefont {M.}~\bibnamefont {Kargarian}},
  \bibinfo {author} {\bibfnamefont {A.}~\bibnamefont {Stern}}, \bibinfo
  {author} {\bibfnamefont {D.}~\bibnamefont {Yue}}, \bibinfo {author}
  {\bibfnamefont {H.}~\bibnamefont {Zhou}}, \bibinfo {author} {\bibfnamefont
  {X.}~\bibnamefont {Jin}}, \bibinfo {author} {\bibfnamefont {V.~M.}\
  \bibnamefont {Galitski}}, \bibinfo {author} {\bibfnamefont {V.~M.}\
  \bibnamefont {Yakovenko}}, \ and\ \bibinfo {author} {\bibfnamefont
  {J.}~\bibnamefont {Xia}},\ }\href@noop {} {\bibfield  {journal} {\bibinfo
  {journal} {Science Advances}\ }\textbf {\bibinfo {volume} {3}},\ \bibinfo
  {pages} {e1602579} (\bibinfo {year} {2017})}\BibitemShut {NoStop}%
\bibitem [{\citenamefont {Beenakker}(2013)}]{Beenakker2013a}%
  \BibitemOpen
  \bibfield  {author} {\bibinfo {author} {\bibfnamefont {C.}~\bibnamefont
  {Beenakker}},\ }\href@noop {} {\bibfield  {journal} {\bibinfo  {journal}
  {Annual Review of Condensed Matter Physics}\ }\textbf {\bibinfo {volume}
  {4}},\ \bibinfo {pages} {113} (\bibinfo {year} {2013})}\BibitemShut {NoStop}%
\bibitem [{\citenamefont {Kwon}\ \emph {et~al.}(2004)\citenamefont {Kwon},
  \citenamefont {Sengupta},\ and\ \citenamefont {Yakovenko}}]{Kwon2004}%
  \BibitemOpen
  \bibfield  {author} {\bibinfo {author} {\bibfnamefont {H.-J.}\ \bibnamefont
  {Kwon}}, \bibinfo {author} {\bibfnamefont {K.}~\bibnamefont {Sengupta}}, \
  and\ \bibinfo {author} {\bibfnamefont {V.~M.}\ \bibnamefont {Yakovenko}},\
  }\href {\doibase 10.1140/epjb/e2004-00066-4} {\bibfield  {journal} {\bibinfo
  {journal} {European Physical Journal B}\ }\textbf {\bibinfo {volume} {37}},\
  \bibinfo {pages} {349} (\bibinfo {year} {2004})}\BibitemShut {NoStop}%
\bibitem [{\citenamefont {Olund}\ and\ \citenamefont {Zhao}(2012)}]{Olund2012}%
  \BibitemOpen
  \bibfield  {author} {\bibinfo {author} {\bibfnamefont {C.~T.}\ \bibnamefont
  {Olund}}\ and\ \bibinfo {author} {\bibfnamefont {E.}~\bibnamefont {Zhao}},\
  }\href {\doibase 10.1103/PhysRevB.86.214515} {\bibfield  {journal} {\bibinfo
  {journal} {Physical Review B}\ }\textbf {\bibinfo {volume} {86}},\ \bibinfo
  {pages} {214515} (\bibinfo {year} {2012})}\BibitemShut {NoStop}%
\bibitem [{\citenamefont {Burset}\ \emph {et~al.}(2014)\citenamefont {Burset},
  \citenamefont {Keidel}, \citenamefont {Tanaka}, \citenamefont {Nagaosa},\
  and\ \citenamefont {Trauzettel}}]{Burset2014}%
  \BibitemOpen
  \bibfield  {author} {\bibinfo {author} {\bibfnamefont {P.}~\bibnamefont
  {Burset}}, \bibinfo {author} {\bibfnamefont {F.}~\bibnamefont {Keidel}},
  \bibinfo {author} {\bibfnamefont {Y.}~\bibnamefont {Tanaka}}, \bibinfo
  {author} {\bibfnamefont {N.}~\bibnamefont {Nagaosa}}, \ and\ \bibinfo
  {author} {\bibfnamefont {B.}~\bibnamefont {Trauzettel}},\ }\href@noop {}
  {\bibfield  {journal} {\bibinfo  {journal} {Physical Review B}\ }\textbf
  {\bibinfo {volume} {90}},\ \bibinfo {pages} {085438} (\bibinfo {year}
  {2014})}\BibitemShut {NoStop}%
\bibitem [{\citenamefont {Ghaemi}\ and\ \citenamefont
  {Nair}(2016)}]{Ghaemi2016}%
  \BibitemOpen
  \bibfield  {author} {\bibinfo {author} {\bibfnamefont {P.}~\bibnamefont
  {Ghaemi}}\ and\ \bibinfo {author} {\bibfnamefont {V.~P.}\ \bibnamefont
  {Nair}},\ }\href {\doibase 10.1103/PhysRevLett.116.037001} {\bibfield
  {journal} {\bibinfo  {journal} {Physical Review Letters}\ }\textbf {\bibinfo
  {volume} {116}},\ \bibinfo {pages} {037001} (\bibinfo {year}
  {2016})}\BibitemShut {NoStop}%
\bibitem [{\citenamefont {Tkachov}(2017)}]{Tkachov2017}%
  \BibitemOpen
  \bibfield  {author} {\bibinfo {author} {\bibfnamefont {G.}~\bibnamefont
  {Tkachov}},\ }\href@noop {} {\bibfield  {journal} {\bibinfo  {journal}
  {Physical Review Letters}\ }\textbf {\bibinfo {volume} {118}},\ \bibinfo
  {pages} {016802} (\bibinfo {year} {2017})}\BibitemShut {NoStop}%
\bibitem [{\citenamefont {Sac{\'{e}}p{\'{e}}}\ \emph
  {et~al.}(2011)\citenamefont {Sac{\'{e}}p{\'{e}}}, \citenamefont {Oostinga},
  \citenamefont {Li}, \citenamefont {Ubaldini}, \citenamefont {Couto},
  \citenamefont {Giannini},\ and\ \citenamefont {Morpurgo}}]{Sacepe2011}%
  \BibitemOpen
  \bibfield  {author} {\bibinfo {author} {\bibfnamefont {B.}~\bibnamefont
  {Sac{\'{e}}p{\'{e}}}}, \bibinfo {author} {\bibfnamefont {J.~B.}\ \bibnamefont
  {Oostinga}}, \bibinfo {author} {\bibfnamefont {J.}~\bibnamefont {Li}},
  \bibinfo {author} {\bibfnamefont {A.}~\bibnamefont {Ubaldini}}, \bibinfo
  {author} {\bibfnamefont {N.~J.}\ \bibnamefont {Couto}}, \bibinfo {author}
  {\bibfnamefont {E.}~\bibnamefont {Giannini}}, \ and\ \bibinfo {author}
  {\bibfnamefont {A.~F.}\ \bibnamefont {Morpurgo}},\ }\href {\doibase
  10.1038/ncomms1586} {\bibfield  {journal} {\bibinfo  {journal} {Nature
  Communications}\ }\textbf {\bibinfo {volume} {2}},\ \bibinfo {pages} {575}
  (\bibinfo {year} {2011})}\BibitemShut {NoStop}%
\bibitem [{\citenamefont {Zhang}\ \emph {et~al.}(2011)\citenamefont {Zhang},
  \citenamefont {Wang}, \citenamefont {DaSilva}, \citenamefont {Lee},
  \citenamefont {Gutierrez}, \citenamefont {Chan}, \citenamefont {Jain},\ and\
  \citenamefont {Samarth}}]{Zhang2011}%
  \BibitemOpen
  \bibfield  {author} {\bibinfo {author} {\bibfnamefont {D.}~\bibnamefont
  {Zhang}}, \bibinfo {author} {\bibfnamefont {J.}~\bibnamefont {Wang}},
  \bibinfo {author} {\bibfnamefont {A.~M.}\ \bibnamefont {DaSilva}}, \bibinfo
  {author} {\bibfnamefont {J.~S.}\ \bibnamefont {Lee}}, \bibinfo {author}
  {\bibfnamefont {H.~R.}\ \bibnamefont {Gutierrez}}, \bibinfo {author}
  {\bibfnamefont {M.~H.~W.}\ \bibnamefont {Chan}}, \bibinfo {author}
  {\bibfnamefont {J.}~\bibnamefont {Jain}}, \ and\ \bibinfo {author}
  {\bibfnamefont {N.}~\bibnamefont {Samarth}},\ }\href {\doibase
  10.1103/PhysRevB.84.165120} {\bibfield  {journal} {\bibinfo  {journal}
  {Physical Review B}\ }\textbf {\bibinfo {volume} {84}},\ \bibinfo {pages}
  {165120} (\bibinfo {year} {2011})}\BibitemShut {NoStop}%
\bibitem [{\citenamefont {Veldhorst}\ \emph {et~al.}(2012)\citenamefont
  {Veldhorst}, \citenamefont {Snelder}, \citenamefont {Hoek}, \citenamefont
  {Gang}, \citenamefont {Guduru}, \citenamefont {Wang}, \citenamefont
  {Zeitler}, \citenamefont {van~der Wiel}, \citenamefont {Golubov},
  \citenamefont {Hilgenkamp},\ and\ \citenamefont {Brinkman}}]{Veldhorst2012}%
  \BibitemOpen
  \bibfield  {author} {\bibinfo {author} {\bibfnamefont {M.}~\bibnamefont
  {Veldhorst}}, \bibinfo {author} {\bibfnamefont {M.}~\bibnamefont {Snelder}},
  \bibinfo {author} {\bibfnamefont {M.}~\bibnamefont {Hoek}}, \bibinfo {author}
  {\bibfnamefont {T.}~\bibnamefont {Gang}}, \bibinfo {author} {\bibfnamefont
  {V.~K.}\ \bibnamefont {Guduru}}, \bibinfo {author} {\bibfnamefont {X.~L.}\
  \bibnamefont {Wang}}, \bibinfo {author} {\bibfnamefont {U.}~\bibnamefont
  {Zeitler}}, \bibinfo {author} {\bibfnamefont {W.~G.}\ \bibnamefont {van~der
  Wiel}}, \bibinfo {author} {\bibfnamefont {A.~A.}\ \bibnamefont {Golubov}},
  \bibinfo {author} {\bibfnamefont {H.}~\bibnamefont {Hilgenkamp}}, \ and\
  \bibinfo {author} {\bibfnamefont {A.}~\bibnamefont {Brinkman}},\ }\href
  {\doibase 10.1038/nmat3255} {\bibfield  {journal} {\bibinfo  {journal}
  {Nature Materials}\ }\textbf {\bibinfo {volume} {11}},\ \bibinfo {pages}
  {417} (\bibinfo {year} {2012})}\BibitemShut {NoStop}%
\bibitem [{\citenamefont {Williams}\ \emph {et~al.}(2012)\citenamefont
  {Williams}, \citenamefont {Bestwick}, \citenamefont {Gallagher},
  \citenamefont {Hong}, \citenamefont {Cui}, \citenamefont {Bleich},
  \citenamefont {Analytis}, \citenamefont {Fisher},\ and\ \citenamefont
  {Goldhaber-Gordon}}]{Williams2012}%
  \BibitemOpen
  \bibfield  {author} {\bibinfo {author} {\bibfnamefont {J.~R.}\ \bibnamefont
  {Williams}}, \bibinfo {author} {\bibfnamefont {A.~J.}\ \bibnamefont
  {Bestwick}}, \bibinfo {author} {\bibfnamefont {P.}~\bibnamefont {Gallagher}},
  \bibinfo {author} {\bibfnamefont {S.~S.}\ \bibnamefont {Hong}}, \bibinfo
  {author} {\bibfnamefont {Y.}~\bibnamefont {Cui}}, \bibinfo {author}
  {\bibfnamefont {A.~S.}\ \bibnamefont {Bleich}}, \bibinfo {author}
  {\bibfnamefont {J.~G.}\ \bibnamefont {Analytis}}, \bibinfo {author}
  {\bibfnamefont {I.~R.}\ \bibnamefont {Fisher}}, \ and\ \bibinfo {author}
  {\bibfnamefont {D.}~\bibnamefont {Goldhaber-Gordon}},\ }\href {\doibase
  10.1103/PhysRevLett.109.056803} {\bibfield  {journal} {\bibinfo  {journal}
  {Physical Review Letters}\ }\textbf {\bibinfo {volume} {109}},\ \bibinfo
  {pages} {056803} (\bibinfo {year} {2012})}\BibitemShut {NoStop}%
\bibitem [{\citenamefont {Qu}\ \emph {et~al.}(2012)\citenamefont {Qu},
  \citenamefont {Yang}, \citenamefont {Shen}, \citenamefont {Ding},
  \citenamefont {Chen}, \citenamefont {Ji}, \citenamefont {Liu}, \citenamefont
  {Fan}, \citenamefont {Jing}, \citenamefont {Yang},\ and\ \citenamefont
  {Lu}}]{Qu2012}%
  \BibitemOpen
  \bibfield  {author} {\bibinfo {author} {\bibfnamefont {F.}~\bibnamefont
  {Qu}}, \bibinfo {author} {\bibfnamefont {F.}~\bibnamefont {Yang}}, \bibinfo
  {author} {\bibfnamefont {J.}~\bibnamefont {Shen}}, \bibinfo {author}
  {\bibfnamefont {Y.}~\bibnamefont {Ding}}, \bibinfo {author} {\bibfnamefont
  {J.}~\bibnamefont {Chen}}, \bibinfo {author} {\bibfnamefont {Z.}~\bibnamefont
  {Ji}}, \bibinfo {author} {\bibfnamefont {G.}~\bibnamefont {Liu}}, \bibinfo
  {author} {\bibfnamefont {J.}~\bibnamefont {Fan}}, \bibinfo {author}
  {\bibfnamefont {X.}~\bibnamefont {Jing}}, \bibinfo {author} {\bibfnamefont
  {C.}~\bibnamefont {Yang}}, \ and\ \bibinfo {author} {\bibfnamefont
  {L.}~\bibnamefont {Lu}},\ }\href {\doibase 10.1038/srep00339} {\bibfield
  {journal} {\bibinfo  {journal} {Scientific Reports}\ }\textbf {\bibinfo
  {volume} {2}},\ \bibinfo {pages} {339} (\bibinfo {year} {2012})}\BibitemShut
  {NoStop}%
\bibitem [{\citenamefont {Yang}\ \emph {et~al.}(2012)\citenamefont {Yang},
  \citenamefont {Qu}, \citenamefont {Shen}, \citenamefont {Ding}, \citenamefont
  {Chen}, \citenamefont {Ji}, \citenamefont {Liu}, \citenamefont {Fan},
  \citenamefont {Yang}, \citenamefont {Fu},\ and\ \citenamefont
  {Lu}}]{Yang2012}%
  \BibitemOpen
  \bibfield  {author} {\bibinfo {author} {\bibfnamefont {F.}~\bibnamefont
  {Yang}}, \bibinfo {author} {\bibfnamefont {F.}~\bibnamefont {Qu}}, \bibinfo
  {author} {\bibfnamefont {J.}~\bibnamefont {Shen}}, \bibinfo {author}
  {\bibfnamefont {Y.}~\bibnamefont {Ding}}, \bibinfo {author} {\bibfnamefont
  {J.}~\bibnamefont {Chen}}, \bibinfo {author} {\bibfnamefont {Z.}~\bibnamefont
  {Ji}}, \bibinfo {author} {\bibfnamefont {G.}~\bibnamefont {Liu}}, \bibinfo
  {author} {\bibfnamefont {J.}~\bibnamefont {Fan}}, \bibinfo {author}
  {\bibfnamefont {C.}~\bibnamefont {Yang}}, \bibinfo {author} {\bibfnamefont
  {L.}~\bibnamefont {Fu}}, \ and\ \bibinfo {author} {\bibfnamefont
  {L.}~\bibnamefont {Lu}},\ }\href {\doibase 10.1103/PhysRevB.86.134504}
  {\bibfield  {journal} {\bibinfo  {journal} {Physical Review B}\ }\textbf
  {\bibinfo {volume} {86}},\ \bibinfo {pages} {134504} (\bibinfo {year}
  {2012})}\BibitemShut {NoStop}%
\bibitem [{\citenamefont {Sochnikov}\ \emph {et~al.}(2013)\citenamefont
  {Sochnikov}, \citenamefont {Bestwick}, \citenamefont {Williams},
  \citenamefont {Lippman}, \citenamefont {Fisher}, \citenamefont
  {Goldhaber-Gordon}, \citenamefont {Kirtley},\ and\ \citenamefont
  {Moler}}]{Sochnikov2013}%
  \BibitemOpen
  \bibfield  {author} {\bibinfo {author} {\bibfnamefont {I.}~\bibnamefont
  {Sochnikov}}, \bibinfo {author} {\bibfnamefont {A.~J.}\ \bibnamefont
  {Bestwick}}, \bibinfo {author} {\bibfnamefont {J.~R.}\ \bibnamefont
  {Williams}}, \bibinfo {author} {\bibfnamefont {T.~M.}\ \bibnamefont
  {Lippman}}, \bibinfo {author} {\bibfnamefont {I.~R.}\ \bibnamefont {Fisher}},
  \bibinfo {author} {\bibfnamefont {D.}~\bibnamefont {Goldhaber-Gordon}},
  \bibinfo {author} {\bibfnamefont {J.~R.}\ \bibnamefont {Kirtley}}, \ and\
  \bibinfo {author} {\bibfnamefont {K.~A.}\ \bibnamefont {Moler}},\ }\href
  {\doibase 10.1021/nl400997k} {\bibfield  {journal} {\bibinfo  {journal} {Nano
  Letters}\ }\textbf {\bibinfo {volume} {13}},\ \bibinfo {pages} {3086}
  (\bibinfo {year} {2013})},\ \bibinfo {note} {pMID: 23795666},\ \Eprint
  {http://arxiv.org/abs/https://doi.org/10.1021/nl400997k}
  {https://doi.org/10.1021/nl400997k} \BibitemShut {NoStop}%
\bibitem [{\citenamefont {Oostinga}\ \emph {et~al.}(2013)\citenamefont
  {Oostinga}, \citenamefont {Maier}, \citenamefont {Sch{\"{u}}ffelgen},
  \citenamefont {Knott}, \citenamefont {Ames}, \citenamefont {Br{\"{u}}ne},
  \citenamefont {Tkachov}, \citenamefont {Buhmann},\ and\ \citenamefont
  {Molenkamp}}]{Oostinga2013}%
  \BibitemOpen
  \bibfield  {author} {\bibinfo {author} {\bibfnamefont {J.~B.}\ \bibnamefont
  {Oostinga}}, \bibinfo {author} {\bibfnamefont {L.}~\bibnamefont {Maier}},
  \bibinfo {author} {\bibfnamefont {P.}~\bibnamefont {Sch{\"{u}}ffelgen}},
  \bibinfo {author} {\bibfnamefont {D.}~\bibnamefont {Knott}}, \bibinfo
  {author} {\bibfnamefont {C.}~\bibnamefont {Ames}}, \bibinfo {author}
  {\bibfnamefont {C.}~\bibnamefont {Br{\"{u}}ne}}, \bibinfo {author}
  {\bibfnamefont {G.}~\bibnamefont {Tkachov}}, \bibinfo {author} {\bibfnamefont
  {H.}~\bibnamefont {Buhmann}}, \ and\ \bibinfo {author} {\bibfnamefont
  {L.~W.}\ \bibnamefont {Molenkamp}},\ }\href {\doibase
  10.1103/PhysRevX.3.021007} {\bibfield  {journal} {\bibinfo  {journal}
  {Physical Review X}\ }\textbf {\bibinfo {volume} {3}},\ \bibinfo {pages}
  {021007} (\bibinfo {year} {2013})}\BibitemShut {NoStop}%
\bibitem [{\citenamefont {Cho}\ \emph {et~al.}(2013)\citenamefont {Cho},
  \citenamefont {Dellabetta}, \citenamefont {Yang}, \citenamefont {Schneeloch},
  \citenamefont {Xu}, \citenamefont {Valla}, \citenamefont {Gu}, \citenamefont
  {Gilbert},\ and\ \citenamefont {Mason}}]{Cho2013}%
  \BibitemOpen
  \bibfield  {author} {\bibinfo {author} {\bibfnamefont {S.}~\bibnamefont
  {Cho}}, \bibinfo {author} {\bibfnamefont {B.}~\bibnamefont {Dellabetta}},
  \bibinfo {author} {\bibfnamefont {A.}~\bibnamefont {Yang}}, \bibinfo {author}
  {\bibfnamefont {J.}~\bibnamefont {Schneeloch}}, \bibinfo {author}
  {\bibfnamefont {Z.}~\bibnamefont {Xu}}, \bibinfo {author} {\bibfnamefont
  {T.}~\bibnamefont {Valla}}, \bibinfo {author} {\bibfnamefont
  {G.}~\bibnamefont {Gu}}, \bibinfo {author} {\bibfnamefont {M.~J.}\
  \bibnamefont {Gilbert}}, \ and\ \bibinfo {author} {\bibfnamefont
  {N.}~\bibnamefont {Mason}},\ }\href {\doibase 10.1038/ncomms2701} {\bibfield
  {journal} {\bibinfo  {journal} {Nature Communications}\ }\textbf {\bibinfo
  {volume} {4}},\ \bibinfo {pages} {1689} (\bibinfo {year} {2013})}\BibitemShut
  {NoStop}%
\bibitem [{\citenamefont {Kurter}\ \emph {et~al.}(2015)\citenamefont {Kurter},
  \citenamefont {Finck}, \citenamefont {Hor},\ and\ \citenamefont {{Van
  Harlingen}}}]{Kurter2013}%
  \BibitemOpen
  \bibfield  {author} {\bibinfo {author} {\bibfnamefont {C.}~\bibnamefont
  {Kurter}}, \bibinfo {author} {\bibfnamefont {A.~D.~K.}\ \bibnamefont
  {Finck}}, \bibinfo {author} {\bibfnamefont {Y.~S.}\ \bibnamefont {Hor}}, \
  and\ \bibinfo {author} {\bibfnamefont {D.~J.}\ \bibnamefont {{Van
  Harlingen}}},\ }\href {\doibase 10.1038/ncomms8130} {\bibfield  {journal}
  {\bibinfo  {journal} {Nature Communications}\ }\textbf {\bibinfo {volume}
  {6}},\ \bibinfo {pages} {7130} (\bibinfo {year} {2015})}\BibitemShut
  {NoStop}%
\bibitem [{\citenamefont {Finck}\ \emph {et~al.}(2014)\citenamefont {Finck},
  \citenamefont {Kurter}, \citenamefont {Hor},\ and\ \citenamefont {{Van
  Harlingen}}}]{Finck2014}%
  \BibitemOpen
  \bibfield  {author} {\bibinfo {author} {\bibfnamefont {A.~D.~K.}\
  \bibnamefont {Finck}}, \bibinfo {author} {\bibfnamefont {C.}~\bibnamefont
  {Kurter}}, \bibinfo {author} {\bibfnamefont {Y.~S.}\ \bibnamefont {Hor}}, \
  and\ \bibinfo {author} {\bibfnamefont {D.~J.}\ \bibnamefont {{Van
  Harlingen}}},\ }\href {\doibase 10.1103/PhysRevX.4.041022} {\bibfield
  {journal} {\bibinfo  {journal} {Physical Review X}\ }\textbf {\bibinfo
  {volume} {4}},\ \bibinfo {pages} {041022} (\bibinfo {year}
  {2014})}\BibitemShut {NoStop}%
\bibitem [{\citenamefont {Lee}\ \emph {et~al.}(2014)\citenamefont {Lee},
  \citenamefont {Lee}, \citenamefont {Park}, \citenamefont {Lee}, \citenamefont
  {Nam}, \citenamefont {Shin}, \citenamefont {Kim},\ and\ \citenamefont
  {Lee}}]{Lee2014a}%
  \BibitemOpen
  \bibfield  {author} {\bibinfo {author} {\bibfnamefont {J.~H.}\ \bibnamefont
  {Lee}}, \bibinfo {author} {\bibfnamefont {G.-H.}\ \bibnamefont {Lee}},
  \bibinfo {author} {\bibfnamefont {J.}~\bibnamefont {Park}}, \bibinfo {author}
  {\bibfnamefont {J.}~\bibnamefont {Lee}}, \bibinfo {author} {\bibfnamefont
  {S.-G.}\ \bibnamefont {Nam}}, \bibinfo {author} {\bibfnamefont {Y.-S.}\
  \bibnamefont {Shin}}, \bibinfo {author} {\bibfnamefont {J.~S.}\ \bibnamefont
  {Kim}}, \ and\ \bibinfo {author} {\bibfnamefont {H.-J.}\ \bibnamefont
  {Lee}},\ }\href {\doibase 10.1021/nl501481b} {\bibfield  {journal} {\bibinfo
  {journal} {Nano Letters}\ }\textbf {\bibinfo {volume} {14}},\ \bibinfo
  {pages} {5029} (\bibinfo {year} {2014})}\BibitemShut {NoStop}%
\bibitem [{\citenamefont {Wiedenmann}\ \emph {et~al.}(2016)\citenamefont
  {Wiedenmann}, \citenamefont {Bocquillon}, \citenamefont {Deacon},
  \citenamefont {Hartinger}, \citenamefont {Herrmann}, \citenamefont
  {Klapwijk}, \citenamefont {Maier}, \citenamefont {Ames}, \citenamefont
  {Br{\"{u}}ne}, \citenamefont {Gould}, \citenamefont {Oiwa}, \citenamefont
  {Ishibashi}, \citenamefont {Tarucha}, \citenamefont {Buhmann},\ and\
  \citenamefont {Molenkamp}}]{Wiedenmann2015}%
  \BibitemOpen
  \bibfield  {author} {\bibinfo {author} {\bibfnamefont {J.}~\bibnamefont
  {Wiedenmann}}, \bibinfo {author} {\bibfnamefont {E.}~\bibnamefont
  {Bocquillon}}, \bibinfo {author} {\bibfnamefont {R.~S.}\ \bibnamefont
  {Deacon}}, \bibinfo {author} {\bibfnamefont {S.}~\bibnamefont {Hartinger}},
  \bibinfo {author} {\bibfnamefont {O.}~\bibnamefont {Herrmann}}, \bibinfo
  {author} {\bibfnamefont {T.~M.}\ \bibnamefont {Klapwijk}}, \bibinfo {author}
  {\bibfnamefont {L.}~\bibnamefont {Maier}}, \bibinfo {author} {\bibfnamefont
  {C.}~\bibnamefont {Ames}}, \bibinfo {author} {\bibfnamefont {C.}~\bibnamefont
  {Br{\"{u}}ne}}, \bibinfo {author} {\bibfnamefont {C.}~\bibnamefont {Gould}},
  \bibinfo {author} {\bibfnamefont {A.}~\bibnamefont {Oiwa}}, \bibinfo {author}
  {\bibfnamefont {K.}~\bibnamefont {Ishibashi}}, \bibinfo {author}
  {\bibfnamefont {S.}~\bibnamefont {Tarucha}}, \bibinfo {author} {\bibfnamefont
  {H.}~\bibnamefont {Buhmann}}, \ and\ \bibinfo {author} {\bibfnamefont
  {L.~W.}\ \bibnamefont {Molenkamp}},\ }\href {\doibase 10.1038/ncomms10303}
  {\bibfield  {journal} {\bibinfo  {journal} {Nature Communications}\ }\textbf
  {\bibinfo {volume} {7}},\ \bibinfo {pages} {10303} (\bibinfo {year}
  {2016})}\BibitemShut {NoStop}%
\bibitem [{\citenamefont {Stehno}\ \emph {et~al.}(2016)\citenamefont {Stehno},
  \citenamefont {Orlyanchik}, \citenamefont {Nugroho}, \citenamefont {Ghaemi},
  \citenamefont {Brahlek}, \citenamefont {Koirala}, \citenamefont {Oh},\ and\
  \citenamefont {{Van Harlingen}}}]{Stehno2016}%
  \BibitemOpen
  \bibfield  {author} {\bibinfo {author} {\bibfnamefont {M.~P.}\ \bibnamefont
  {Stehno}}, \bibinfo {author} {\bibfnamefont {V.}~\bibnamefont {Orlyanchik}},
  \bibinfo {author} {\bibfnamefont {C.~D.}\ \bibnamefont {Nugroho}}, \bibinfo
  {author} {\bibfnamefont {P.}~\bibnamefont {Ghaemi}}, \bibinfo {author}
  {\bibfnamefont {M.}~\bibnamefont {Brahlek}}, \bibinfo {author} {\bibfnamefont
  {N.}~\bibnamefont {Koirala}}, \bibinfo {author} {\bibfnamefont
  {S.}~\bibnamefont {Oh}}, \ and\ \bibinfo {author} {\bibfnamefont {D.~J.}\
  \bibnamefont {{Van Harlingen}}},\ }\href
  {https://journals.aps.org/prb/pdf/10.1103/PhysRevB.93.035307} {\bibfield
  {journal} {\bibinfo  {journal} {Physical Review B}\ }\textbf {\bibinfo
  {volume} {93}},\ \bibinfo {pages} {035307} (\bibinfo {year}
  {2016})}\BibitemShut {NoStop}%
\bibitem [{\citenamefont {Bocquillon}\ \emph {et~al.}(2017)\citenamefont
  {Bocquillon}, \citenamefont {Deacon}, \citenamefont {Wiedenmann},
  \citenamefont {Leubner}, \citenamefont {Klapwijk}, \citenamefont
  {Br{\"{u}}ne}, \citenamefont {Ishibashi}, \citenamefont {Buhmann},\ and\
  \citenamefont {Molenkamp}}]{Bocquillon2017}%
  \BibitemOpen
  \bibfield  {author} {\bibinfo {author} {\bibfnamefont {E.}~\bibnamefont
  {Bocquillon}}, \bibinfo {author} {\bibfnamefont {R.~S.}\ \bibnamefont
  {Deacon}}, \bibinfo {author} {\bibfnamefont {J.}~\bibnamefont {Wiedenmann}},
  \bibinfo {author} {\bibfnamefont {P.}~\bibnamefont {Leubner}}, \bibinfo
  {author} {\bibfnamefont {T.~M.}\ \bibnamefont {Klapwijk}}, \bibinfo {author}
  {\bibfnamefont {C.}~\bibnamefont {Br{\"{u}}ne}}, \bibinfo {author}
  {\bibfnamefont {K.}~\bibnamefont {Ishibashi}}, \bibinfo {author}
  {\bibfnamefont {H.}~\bibnamefont {Buhmann}}, \ and\ \bibinfo {author}
  {\bibfnamefont {L.~W.}\ \bibnamefont {Molenkamp}},\ }\href {\doibase
  10.1038/nnano.2016.159} {\bibfield  {journal} {\bibinfo  {journal} {Nature
  Nanotechnology}\ }\textbf {\bibinfo {volume} {12}},\ \bibinfo {pages} {137}
  (\bibinfo {year} {2017})}\BibitemShut {NoStop}%
\bibitem [{\citenamefont {Jauregui}\ \emph {et~al.}(2018)\citenamefont
  {Jauregui}, \citenamefont {Kayyalha}, \citenamefont {Kazakov}, \citenamefont
  {Miotkowski}, \citenamefont {Rokhinson},\ and\ \citenamefont
  {Chen}}]{Jauregui2018}%
  \BibitemOpen
  \bibfield  {author} {\bibinfo {author} {\bibfnamefont {L.~A.}\ \bibnamefont
  {Jauregui}}, \bibinfo {author} {\bibfnamefont {M.}~\bibnamefont {Kayyalha}},
  \bibinfo {author} {\bibfnamefont {A.}~\bibnamefont {Kazakov}}, \bibinfo
  {author} {\bibfnamefont {I.}~\bibnamefont {Miotkowski}}, \bibinfo {author}
  {\bibfnamefont {L.~P.}\ \bibnamefont {Rokhinson}}, \ and\ \bibinfo {author}
  {\bibfnamefont {Y.~P.}\ \bibnamefont {Chen}},\ }\href@noop {} {\bibfield
  {journal} {\bibinfo  {journal} {Applied Physics Letters}\ }\textbf {\bibinfo
  {volume} {112}},\ \bibinfo {pages} {093105} (\bibinfo {year}
  {2018})}\BibitemShut {NoStop}%
\bibitem [{\citenamefont {Xu}\ \emph {et~al.}(2014)\citenamefont {Xu},
  \citenamefont {Miotkowski}, \citenamefont {Liu}, \citenamefont {Tian},
  \citenamefont {Nam}, \citenamefont {Alidoust}, \citenamefont {Hu},
  \citenamefont {Shih}, \citenamefont {Hasan},\ and\ \citenamefont
  {Chen}}]{Xu2014}%
  \BibitemOpen
  \bibfield  {author} {\bibinfo {author} {\bibfnamefont {Y.}~\bibnamefont
  {Xu}}, \bibinfo {author} {\bibfnamefont {I.}~\bibnamefont {Miotkowski}},
  \bibinfo {author} {\bibfnamefont {C.}~\bibnamefont {Liu}}, \bibinfo {author}
  {\bibfnamefont {J.}~\bibnamefont {Tian}}, \bibinfo {author} {\bibfnamefont
  {H.}~\bibnamefont {Nam}}, \bibinfo {author} {\bibfnamefont {N.}~\bibnamefont
  {Alidoust}}, \bibinfo {author} {\bibfnamefont {J.}~\bibnamefont {Hu}},
  \bibinfo {author} {\bibfnamefont {C.-K.}\ \bibnamefont {Shih}}, \bibinfo
  {author} {\bibfnamefont {M.~Z.}\ \bibnamefont {Hasan}}, \ and\ \bibinfo
  {author} {\bibfnamefont {Y.~P.}\ \bibnamefont {Chen}},\ }\href {\doibase
  10.1038/nphys3140} {\bibfield  {journal} {\bibinfo  {journal} {Nature
  Physics}\ }\textbf {\bibinfo {volume} {10}},\ \bibinfo {pages} {956}
  (\bibinfo {year} {2014})}\BibitemShut {NoStop}%
\bibitem [{\citenamefont {Xu}\ \emph {et~al.}(2016)\citenamefont {Xu},
  \citenamefont {Miotkowski},\ and\ \citenamefont {Chen}}]{Xu2016}%
  \BibitemOpen
  \bibfield  {author} {\bibinfo {author} {\bibfnamefont {Y.}~\bibnamefont
  {Xu}}, \bibinfo {author} {\bibfnamefont {I.}~\bibnamefont {Miotkowski}}, \
  and\ \bibinfo {author} {\bibfnamefont {Y.~P.}\ \bibnamefont {Chen}},\ }\href
  {\doibase 10.1038/ncomms11434} {\bibfield  {journal} {\bibinfo  {journal}
  {Nature Communications}\ }\textbf {\bibinfo {volume} {7}},\ \bibinfo {pages}
  {11434} (\bibinfo {year} {2016})}\BibitemShut {NoStop}%
\bibitem [{\citenamefont {Hong}\ \emph {et~al.}(2014)\citenamefont {Hong},
  \citenamefont {Zhang}, \citenamefont {Cha}, \citenamefont {Qi},\ and\
  \citenamefont {Cui}}]{Hong2014}%
  \BibitemOpen
  \bibfield  {author} {\bibinfo {author} {\bibfnamefont {S.~S.}\ \bibnamefont
  {Hong}}, \bibinfo {author} {\bibfnamefont {Y.}~\bibnamefont {Zhang}},
  \bibinfo {author} {\bibfnamefont {J.~J.}\ \bibnamefont {Cha}}, \bibinfo
  {author} {\bibfnamefont {X.-L.}\ \bibnamefont {Qi}}, \ and\ \bibinfo {author}
  {\bibfnamefont {Y.}~\bibnamefont {Cui}},\ }\href@noop {} {\bibfield
  {journal} {\bibinfo  {journal} {Nano Letters}\ }\textbf {\bibinfo {volume}
  {14}},\ \bibinfo {pages} {2815} (\bibinfo {year} {2014})}\BibitemShut
  {NoStop}%
\bibitem [{\citenamefont {Cho}\ \emph {et~al.}(2015)\citenamefont {Cho},
  \citenamefont {Dellabetta}, \citenamefont {Zhong}, \citenamefont
  {Schneeloch}, \citenamefont {Liu}, \citenamefont {Gu}, \citenamefont
  {Gilbert},\ and\ \citenamefont {Mason}}]{Cho2015}%
  \BibitemOpen
  \bibfield  {author} {\bibinfo {author} {\bibfnamefont {S.}~\bibnamefont
  {Cho}}, \bibinfo {author} {\bibfnamefont {B.}~\bibnamefont {Dellabetta}},
  \bibinfo {author} {\bibfnamefont {R.}~\bibnamefont {Zhong}}, \bibinfo
  {author} {\bibfnamefont {J.}~\bibnamefont {Schneeloch}}, \bibinfo {author}
  {\bibfnamefont {T.}~\bibnamefont {Liu}}, \bibinfo {author} {\bibfnamefont
  {G.}~\bibnamefont {Gu}}, \bibinfo {author} {\bibfnamefont {M.~J.}\
  \bibnamefont {Gilbert}}, \ and\ \bibinfo {author} {\bibfnamefont
  {N.}~\bibnamefont {Mason}},\ }\href@noop {} {\bibfield  {journal} {\bibinfo
  {journal} {Nature Communications}\ }\textbf {\bibinfo {volume} {6}},\
  \bibinfo {pages} {7634} (\bibinfo {year} {2015})}\BibitemShut {NoStop}%
\bibitem [{\citenamefont {Jauregui}\ \emph {et~al.}(2016)\citenamefont
  {Jauregui}, \citenamefont {Pettes}, \citenamefont {Rokhinson}, \citenamefont
  {Shi},\ and\ \citenamefont {Chen}}]{Jauregui2016}%
  \BibitemOpen
  \bibfield  {author} {\bibinfo {author} {\bibfnamefont {L.~A.}\ \bibnamefont
  {Jauregui}}, \bibinfo {author} {\bibfnamefont {M.~T.}\ \bibnamefont
  {Pettes}}, \bibinfo {author} {\bibfnamefont {L.~P.}\ \bibnamefont
  {Rokhinson}}, \bibinfo {author} {\bibfnamefont {L.}~\bibnamefont {Shi}}, \
  and\ \bibinfo {author} {\bibfnamefont {Y.~P.}\ \bibnamefont {Chen}},\ }\href
  {\doibase 10.1038/nnano.2015.293} {\bibfield  {journal} {\bibinfo  {journal}
  {Nature Nanotechnology}\ }\textbf {\bibinfo {volume} {11}},\ \bibinfo {pages}
  {345} (\bibinfo {year} {2016})}\BibitemShut {NoStop}%
\bibitem [{\citenamefont {Cook}\ and\ \citenamefont {Franz}(2011)}]{Cook2011}%
  \BibitemOpen
  \bibfield  {author} {\bibinfo {author} {\bibfnamefont {A.}~\bibnamefont
  {Cook}}\ and\ \bibinfo {author} {\bibfnamefont {M.}~\bibnamefont {Franz}},\
  }\href {\doibase 10.1103/PhysRevB.84.201105} {\bibfield  {journal} {\bibinfo
  {journal} {Physical Review B}\ }\textbf {\bibinfo {volume} {84}},\ \bibinfo
  {pages} {201105} (\bibinfo {year} {2011})}\BibitemShut {NoStop}%
\bibitem [{\citenamefont {Ilan}\ \emph {et~al.}(2014)\citenamefont {Ilan},
  \citenamefont {Bardarson}, \citenamefont {Sim},\ and\ \citenamefont
  {Moore}}]{Ilan2014}%
  \BibitemOpen
  \bibfield  {author} {\bibinfo {author} {\bibfnamefont {R.}~\bibnamefont
  {Ilan}}, \bibinfo {author} {\bibfnamefont {J.~H.}\ \bibnamefont {Bardarson}},
  \bibinfo {author} {\bibfnamefont {H.~S.}\ \bibnamefont {Sim}}, \ and\
  \bibinfo {author} {\bibfnamefont {J.~E.}\ \bibnamefont {Moore}},\ }\href@noop
  {} {\bibfield  {journal} {\bibinfo  {journal} {New Journal of Physics}\
  }\textbf {\bibinfo {volume} {16}},\ \bibinfo {pages} {053007} (\bibinfo
  {year} {2014})}\BibitemShut {NoStop}%
\bibitem [{SIP()}]{SIPRL2017}%
  \BibitemOpen
  \href@noop {} {}\bibinfo {note} {See Supplemental Material at ,for further
  details regarding the sample parameters, temperature dependence of the
  critical current, and the measurement of the current-phase
  relation}\BibitemShut {NoStop}%
\bibitem [{\citenamefont {Dubos}\ \emph {et~al.}(2001)\citenamefont {Dubos},
  \citenamefont {Courtois}, \citenamefont {Pannetier}, \citenamefont {Wilhelm},
  \citenamefont {Zaikin},\ and\ \citenamefont {Schon}}]{Dubos2001}%
  \BibitemOpen
  \bibfield  {author} {\bibinfo {author} {\bibfnamefont {P.}~\bibnamefont
  {Dubos}}, \bibinfo {author} {\bibfnamefont {H.}~\bibnamefont {Courtois}},
  \bibinfo {author} {\bibfnamefont {B.}~\bibnamefont {Pannetier}}, \bibinfo
  {author} {\bibfnamefont {F.~K.}\ \bibnamefont {Wilhelm}}, \bibinfo {author}
  {\bibfnamefont {A.~D.}\ \bibnamefont {Zaikin}}, \ and\ \bibinfo {author}
  {\bibfnamefont {G.}~\bibnamefont {Schon}},\ }\href {\doibase
  10.1103/PhysRevB.63.064502} {\bibfield  {journal} {\bibinfo  {journal}
  {Physical Review B}\ }\textbf {\bibinfo {volume} {63}},\ \bibinfo {pages}
  {064502} (\bibinfo {year} {2001})}\BibitemShut {NoStop}%
\bibitem [{\citenamefont {Golubov}\ \emph {et~al.}(2004)\citenamefont
  {Golubov}, \citenamefont {Kupriyanov},\ and\ \citenamefont
  {Il'ichev}}]{Golubov2004}%
  \BibitemOpen
  \bibfield  {author} {\bibinfo {author} {\bibfnamefont {A.~A.}\ \bibnamefont
  {Golubov}}, \bibinfo {author} {\bibfnamefont {M.~Y.}\ \bibnamefont
  {Kupriyanov}}, \ and\ \bibinfo {author} {\bibfnamefont {E.}~\bibnamefont
  {Il'ichev}},\ }\href {\doibase 10.1103/RevModPhys.76.411} {\bibfield
  {journal} {\bibinfo  {journal} {Reviews of Modern Physics}\ }\textbf
  {\bibinfo {volume} {76}},\ \bibinfo {pages} {411} (\bibinfo {year}
  {2004})}\BibitemShut {NoStop}%
\bibitem [{\citenamefont {Angers}\ \emph {et~al.}(2008)\citenamefont {Angers},
  \citenamefont {Chiodi}, \citenamefont {Montambaux}, \citenamefont {Ferrier},
  \citenamefont {Gu{\'e}ron}, \citenamefont {Bouchiat},\ and\ \citenamefont
  {Cuevas}}]{Angers2008}%
  \BibitemOpen
  \bibfield  {author} {\bibinfo {author} {\bibfnamefont {L.}~\bibnamefont
  {Angers}}, \bibinfo {author} {\bibfnamefont {F.}~\bibnamefont {Chiodi}},
  \bibinfo {author} {\bibfnamefont {G.}~\bibnamefont {Montambaux}}, \bibinfo
  {author} {\bibfnamefont {M.}~\bibnamefont {Ferrier}}, \bibinfo {author}
  {\bibfnamefont {S.}~\bibnamefont {Gu{\'e}ron}}, \bibinfo {author}
  {\bibfnamefont {H.}~\bibnamefont {Bouchiat}}, \ and\ \bibinfo {author}
  {\bibfnamefont {J.~C.}\ \bibnamefont {Cuevas}},\ }\href@noop {} {\bibfield
  {journal} {\bibinfo  {journal} {Physical Review B}\ }\textbf {\bibinfo
  {volume} {77}},\ \bibinfo {pages} {165408} (\bibinfo {year}
  {2008})}\BibitemShut {NoStop}%
\bibitem [{\citenamefont {Ke}\ \emph {et~al.}(2016)\citenamefont {Ke},
  \citenamefont {Borzenets}, \citenamefont {Draelos}, \citenamefont {Amet},
  \citenamefont {Bomze}, \citenamefont {Jones}, \citenamefont {Craciun},
  \citenamefont {Russo}, \citenamefont {Yamamoto}, \citenamefont {Tarucha}
  \emph {et~al.}}]{Ke2016}%
  \BibitemOpen
  \bibfield  {author} {\bibinfo {author} {\bibfnamefont {C.~T.}\ \bibnamefont
  {Ke}}, \bibinfo {author} {\bibfnamefont {I.~V.}\ \bibnamefont {Borzenets}},
  \bibinfo {author} {\bibfnamefont {A.~W.}\ \bibnamefont {Draelos}}, \bibinfo
  {author} {\bibfnamefont {F.}~\bibnamefont {Amet}}, \bibinfo {author}
  {\bibfnamefont {Y.}~\bibnamefont {Bomze}}, \bibinfo {author} {\bibfnamefont
  {G.}~\bibnamefont {Jones}}, \bibinfo {author} {\bibfnamefont
  {M.}~\bibnamefont {Craciun}}, \bibinfo {author} {\bibfnamefont
  {S.}~\bibnamefont {Russo}}, \bibinfo {author} {\bibfnamefont
  {M.}~\bibnamefont {Yamamoto}}, \bibinfo {author} {\bibfnamefont
  {S.}~\bibnamefont {Tarucha}},  \emph {et~al.},\ }\href@noop {} {\bibfield
  {journal} {\bibinfo  {journal} {Nano Letters}\ }\textbf {\bibinfo {volume}
  {16}},\ \bibinfo {pages} {4788} (\bibinfo {year} {2016})}\BibitemShut
  {NoStop}%
\bibitem [{\citenamefont {Borzenets}\ \emph {et~al.}(2016)\citenamefont
  {Borzenets}, \citenamefont {Amet}, \citenamefont {Ke}, \citenamefont
  {Draelos}, \citenamefont {Wei}, \citenamefont {Seredinski}, \citenamefont
  {Watanabe}, \citenamefont {Taniguchi}, \citenamefont {Bomze}, \citenamefont
  {Yamamoto} \emph {et~al.}}]{Borzenets2016}%
  \BibitemOpen
  \bibfield  {author} {\bibinfo {author} {\bibfnamefont {I.}~\bibnamefont
  {Borzenets}}, \bibinfo {author} {\bibfnamefont {F.}~\bibnamefont {Amet}},
  \bibinfo {author} {\bibfnamefont {C.}~\bibnamefont {Ke}}, \bibinfo {author}
  {\bibfnamefont {A.}~\bibnamefont {Draelos}}, \bibinfo {author} {\bibfnamefont
  {M.}~\bibnamefont {Wei}}, \bibinfo {author} {\bibfnamefont {A.}~\bibnamefont
  {Seredinski}}, \bibinfo {author} {\bibfnamefont {K.}~\bibnamefont
  {Watanabe}}, \bibinfo {author} {\bibfnamefont {T.}~\bibnamefont {Taniguchi}},
  \bibinfo {author} {\bibfnamefont {Y.}~\bibnamefont {Bomze}}, \bibinfo
  {author} {\bibfnamefont {M.}~\bibnamefont {Yamamoto}},  \emph {et~al.},\
  }\href@noop {} {\bibfield  {journal} {\bibinfo  {journal} {Physical Review
  Letters}\ }\textbf {\bibinfo {volume} {117}},\ \bibinfo {pages} {237002}
  (\bibinfo {year} {2016})}\BibitemShut {NoStop}%
\bibitem [{\citenamefont {Zhang}\ and\ \citenamefont
  {Vishwanath}(2010)}]{Zhang2010}%
  \BibitemOpen
  \bibfield  {author} {\bibinfo {author} {\bibfnamefont {Y.}~\bibnamefont
  {Zhang}}\ and\ \bibinfo {author} {\bibfnamefont {A.}~\bibnamefont
  {Vishwanath}},\ }\href@noop {} {\bibfield  {journal} {\bibinfo  {journal}
  {Physical Review Letters}\ }\textbf {\bibinfo {volume} {105}},\ \bibinfo
  {pages} {206601} (\bibinfo {year} {2010})}\BibitemShut {NoStop}%
\bibitem [{\citenamefont {Bardarson}\ \emph {et~al.}(2010)\citenamefont
  {Bardarson}, \citenamefont {Brouwer},\ and\ \citenamefont
  {Moore}}]{Bardarson2010}%
  \BibitemOpen
  \bibfield  {author} {\bibinfo {author} {\bibfnamefont {J.~H.}\ \bibnamefont
  {Bardarson}}, \bibinfo {author} {\bibfnamefont {P.~W.}\ \bibnamefont
  {Brouwer}}, \ and\ \bibinfo {author} {\bibfnamefont {J.~E.}\ \bibnamefont
  {Moore}},\ }\href {\doibase 10.1103/PhysRevLett.105.156803} {\bibfield
  {journal} {\bibinfo  {journal} {Physical Review Letters}\ }\textbf {\bibinfo
  {volume} {105}},\ \bibinfo {pages} {156803} (\bibinfo {year}
  {2010})}\BibitemShut {NoStop}%
\bibitem [{\citenamefont {Tinkham}(1996)}]{Tinkham2004}%
  \BibitemOpen
  \bibfield  {author} {\bibinfo {author} {\bibfnamefont {M.}~\bibnamefont
  {Tinkham}},\ }\href@noop {} {\emph {\bibinfo {title} {{Introduction to
  superconductivity}}}}\ (\bibinfo  {publisher} {McGrawHill, New York},\
  \bibinfo {year} {1996})\ p.\ \bibinfo {pages} {454}\BibitemShut {NoStop}%
\bibitem [{\citenamefont {Bardeen}\ and\ \citenamefont
  {Johnson}(1972)}]{Bardeen1972}%
  \BibitemOpen
  \bibfield  {author} {\bibinfo {author} {\bibfnamefont {J.}~\bibnamefont
  {Bardeen}}\ and\ \bibinfo {author} {\bibfnamefont {J.~L.}\ \bibnamefont
  {Johnson}},\ }\href@noop {} {\bibfield  {journal} {\bibinfo  {journal}
  {Physical Review B}\ }\textbf {\bibinfo {volume} {5}},\ \bibinfo {pages} {72}
  (\bibinfo {year} {1972})}\BibitemShut {NoStop}%
\bibitem [{\citenamefont {Svidzinsky}\ \emph {et~al.}(1973)\citenamefont
  {Svidzinsky}, \citenamefont {Antsygina},\ and\ \citenamefont
  {Bratus}}]{Svidzinsky1973}%
  \BibitemOpen
  \bibfield  {author} {\bibinfo {author} {\bibfnamefont {A.}~\bibnamefont
  {Svidzinsky}}, \bibinfo {author} {\bibfnamefont {T.}~\bibnamefont
  {Antsygina}}, \ and\ \bibinfo {author} {\bibfnamefont {E.~N.}\ \bibnamefont
  {Bratus}},\ }\href@noop {} {\bibfield  {journal} {\bibinfo  {journal}
  {Journal of Low Temperature Physics}\ }\textbf {\bibinfo {volume} {10}},\
  \bibinfo {pages} {131} (\bibinfo {year} {1973})}\BibitemShut {NoStop}%
\bibitem [{\citenamefont {Bagwell}(1992)}]{Bagwell1992}%
  \BibitemOpen
  \bibfield  {author} {\bibinfo {author} {\bibfnamefont {P.~F.}\ \bibnamefont
  {Bagwell}},\ }\href@noop {} {\bibfield  {journal} {\bibinfo  {journal}
  {Physical Review B}\ }\textbf {\bibinfo {volume} {46}},\ \bibinfo {pages}
  {12573} (\bibinfo {year} {1992})}\BibitemShut {NoStop}%
\bibitem [{\citenamefont {{Della Rocca}}\ \emph {et~al.}(2007)\citenamefont
  {{Della Rocca}}, \citenamefont {Chauvin}, \citenamefont {Huard},
  \citenamefont {Pothier}, \citenamefont {Esteve},\ and\ \citenamefont
  {Urbina}}]{DellaRocca2007}%
  \BibitemOpen
  \bibfield  {author} {\bibinfo {author} {\bibfnamefont {M.~L.}\ \bibnamefont
  {{Della Rocca}}}, \bibinfo {author} {\bibfnamefont {M.}~\bibnamefont
  {Chauvin}}, \bibinfo {author} {\bibfnamefont {B.}~\bibnamefont {Huard}},
  \bibinfo {author} {\bibfnamefont {H.}~\bibnamefont {Pothier}}, \bibinfo
  {author} {\bibfnamefont {D.}~\bibnamefont {Esteve}}, \ and\ \bibinfo {author}
  {\bibfnamefont {C.}~\bibnamefont {Urbina}},\ }\href {\doibase
  10.1103/PhysRevLett.99.127005} {\bibfield  {journal} {\bibinfo  {journal}
  {Physical Review Letters}\ }\textbf {\bibinfo {volume} {99}},\ \bibinfo
  {pages} {127005} (\bibinfo {year} {2007})}\BibitemShut {NoStop}%
\bibitem [{\citenamefont {Zgirski}\ \emph {et~al.}(2011)\citenamefont
  {Zgirski}, \citenamefont {Bretheau}, \citenamefont {{Le Masne}},
  \citenamefont {Pothier}, \citenamefont {Esteve},\ and\ \citenamefont
  {Urbina}}]{Zgirski2011}%
  \BibitemOpen
  \bibfield  {author} {\bibinfo {author} {\bibfnamefont {M.}~\bibnamefont
  {Zgirski}}, \bibinfo {author} {\bibfnamefont {L.}~\bibnamefont {Bretheau}},
  \bibinfo {author} {\bibfnamefont {Q.}~\bibnamefont {{Le Masne}}}, \bibinfo
  {author} {\bibfnamefont {H.}~\bibnamefont {Pothier}}, \bibinfo {author}
  {\bibfnamefont {D.}~\bibnamefont {Esteve}}, \ and\ \bibinfo {author}
  {\bibfnamefont {C.}~\bibnamefont {Urbina}},\ }\href {\doibase
  10.1103/PhysRevLett.106.257003} {\bibfield  {journal} {\bibinfo  {journal}
  {Physical Review Letters}\ }\textbf {\bibinfo {volume} {106}},\ \bibinfo
  {pages} {257003} (\bibinfo {year} {2011})}\BibitemShut {NoStop}%
\end{thebibliography}%

\end{document}